\documentclass[twocolumn,showpacs,preprintnumbers,amsmath,amssymb,amsmath]{revtex4-1}

\usepackage{graphicx}
\usepackage{dcolumn}
\usepackage{bm}
\usepackage{hyperref}
\usepackage{epsf}
\usepackage{float} 
\newcommand{\eq}[1]{(\ref{#1})}
\newcommand{\la}{\label}
\newcommand{\eea}{\end{eqnarray}}
\newcommand{\beq}{\begin{equation}}
\newcommand{\eeq}{\end{equation}}
\newcommand{\be}{\begin{equation}}
\newcommand{\ee}{\end{equation}}

\newcommand{\p}{\partial}

\def\XXint#1#2#3{{\setbox0=\hbox{$#1{#2#3}{\int}$ }
\vcenter{\hbox{$#2#3$ }}\kern-.5\wd0}}

\usepackage{amsmath}
\usepackage{amsfonts}
\usepackage{varioref}
\usepackage{amssymb}

\usepackage{color}

\begin{document}

\title{Anomalous Hydrodynamics of Two-Dimensional Vortex Fluid}

\author{Paul Wiegmann}
 \affiliation{ Department of Physics, University of Chicago, 929 57th St, Chicago, IL 60637, USA}
 \author{Alexander G.~Abanov}
\affiliation{Department of Physics and Astronomy and Simons Center for Geometry and Physics,
Stony Brook University,  Stony Brook, NY 11794, USA}

\date{\today}
%

\date{\today}

\begin{abstract}
Turbulent flows of incompressible liquid in two dimensions are comprised of dense systems of vortices. 
 Such  system of vortices  can be treated as a fluid and itself could be described  in terms of hydrodynamics.  We develop the hydrodynamics of the vortex fluid. This hydrodynamics  captures characteristics of fluid flows averaged over fast circulations  in the inter-vortex space. The hydrodynamics of the vortex fluid features the 
 anomalous stress absent in Euler's hydrodynamics. The anomalous  stress yields a number of  interesting effects. Some of them are: a deflection of stream lines, a correction to the Bernoulli law, accumulation of vortices in regions with high curvature in the curved space. The origin of the anomalous stresses is a divergence of inter-vortex interactions at the micro-scale which manifest at the macro-scale. We obtain the  hydrodynamics
of the vortex fluid from the Kirchhoff equations for dynamics of point-like vortices. 
\end{abstract}

                           
\maketitle

\paragraph*{1. Introduction and main results.}    
Turbulent flows of two dimensional   incompressible fluids consist of numerous vortices found essentially at all scales. In many physical realizations larger vortices can be considered as a collective motion of smaller minimal eddies which can not be fragmented further. Fluids  with lower bound of vortex circulations could be both quantum and classical. Examples of quantum fluids include superfluid Helium \cite{Onsager,Feynman, Khalatnikov},  superconductors, cold atomic systems (BEC) \cite{BEC} and electronic liquids in the Fractional Quantum Hall regime \cite{W2},  models of classical turbulence with quantized vortices, sometimes referred to as  models of the quantum turbulence (see e.g. \cite{QT}). The lower bound of vortex  circulation \(2\pi\gamma\)  in quantum fluids is defined by the Planck constant.
In classical fluids the lower bound for vortex circulation can occur in turbulent flows in the regime of the inverse cascade \cite{1980-KraichnanMontgomery}. There it is determined by the injection scale.

In the regime of interest for this work the minimal eddies are spatially separated by a typical scale \(\ell\) which is larger than the size of vortex cores but much smaller than the system size. There is no dissipation on these scales. In such regime the fluid motion can be separated into the fast motion with
typical velocity \(\gamma/\ell\) describing the rotation of the fluid around  vortex
cores and the  slow motion of vortices themselves. Description of large  structures in fluid dynamics requires averaging over fast motions. The averaging essentially eliminates fast motion reducing the fluid dynamics to the hydrodynamics  of the vortex matter. 
Thus, on time scales larger than \(\ell^2/\gamma \) minimal eddies should be treated as a (secondary) dissipation-free fluid. 

The idea to treat vortices as a macroscopical  system goes back to seminal works of Onsager \cite{Onsager}. Onsager suggested to describe statistical properties of turbulent flows in 2D incompressible fluid  by Gibbsian equilibrium statistical mechanics of a gas of point-like vortices  \cite{Onsager}.   In this paper we address the hydrodynamics of the vortex fluid \cite{comment1}. 
Such hydrodynamics  is  a suitable platform   to study  the large scale and long time characteristics
of flows.

The hydrodynamics of the vortex fluid can be obtained as a result of averaging over fast motion in parcels centered around vortices. 
The   average  velocity of each parcel   is approximately equal to the velocity of the vortex itself. Therefore, positions of vortices, or Lagrangian coordinates of the vortex fluid are the   ``slow variables" describing the motion of the fluid.   The goal of this paper is to derive the  Euler description of the vortex fluid.

We focus on the hydrodynamics of  the \emph{chiral
flow} where circulations of all  vortices are of the same sign. Chiral flows occur in rotating
fluids, and also, emerge as a  spontaneous formation of like-circulation stable vortex
clusters predicted Onsager \cite{Onsager}, and commonly  observed in 2D fluids. In this paper we assume that a chiral cluster has already been formed and focus on the vortex dynamics within the cluster.

Our main result is that  the  hydrodynamics of the chiral vortex flow  is anomalous.
Anomalous terms in  hydrodynamics   are conservative  pseudo-tensor parts in the stress tensors \cite{AF}. They are invariant  under  translations and proper rotations but change sign under an orientation reversing coordinates transformation. In two dimensions the anomalous term is linear in velocity. In Cartesian coordinates it reads 
\begin{align}
  \la{2}&
        \tau_{xy} = \tau_{yx} =\eta(\nabla_xv_x - \nabla_yv_y)\,,
 \\
        &\tau_{xx}=-\tau_{yy}=-\eta(\nabla_xv_y+\nabla_yv_x)\,.
  \nonumber
\end{align}
where $\bm{v}=(v_x,v_y)$ is the (Eulerian) velocity of the vortex flow  and \(\eta\) is the anomalous kinetic coefficient.

The anomalous stress is accompanied by the anomalous  addition to the pressure of the vortex fluid 
\begin{align}
 \la{p}
 	\mathfrak{p}=p+ 2\eta|\omega|+2\eta^2\frac{\Delta\sqrt{|\omega|}}{\sqrt{|\omega|}}\,,
\end{align}
 where $p$ is the  pressure and $\omega$ is the vorticity  of the  fluid.   We show that, the  anomalous kinetic coefficient \(\eta\)  is the quarter   of the
minimal vortex strength
\begin{eqnarray}\la{eta}
        \eta=\frac{\gamma}{4} \,.
\end{eqnarray}Euler's incompressible inviscid fluid does not exert a force acting  on a shear flow. The vortex fluid does.  The anomalous  force  exerted by
the  flow on the line element of the vortex fluid  
acts normally to the vortex  shear flow.   It produces neither work nor dissipation.

As a result of anomalous force the stream lines of the vortex flow are no
longer aligned to the stream lines of the liquid flow. This misalignment causes corrections
to the Bernoulli law.

Anomalous forces are revealed in the presence of boundaries and on curved manifolds. In this paper we briefly discuss the latter. On a curved manifold anomalous forces cause vortices to flow toward regions of positive curvature  at the expense of the deficit of vortices in regions of negative curvature. This effect gives, yet another definition of the anomalous kinetic coefficient: a response of the vorticity of the flow $\omega$ to a variation of the (scalar) curvature $R$
\be\eta=\frac{\delta\omega}{\delta R}.\la{48}\ee In the paper we discuss  few  consequences of the anomalous term.

Through the paper we use the complex coordinates \(z=x+i y,\,\p=\frac 12(\p_x-i \p_y)\),
and the  complex   vector components
$u=u_x-i  u_y$. 
  We use subscripts $a,b$ to denote the Cartesian components of vectors and subscripts $i,j,k$ for particle labels. We also use the complex
components for symmetric tensors \(\tau=\tau_{xx}-\tau_{yy}-2i 
\tau_{xy},\;\tau_{z\bar z}=\tau_{xx}+\tau_{yy}\).
In complex notations the components of the anomalous stress read
\begin{align}
	\tau=-4i \eta \p v,\quad\tau_{z\bar z}=4i\eta  \bar \p v.\la{40}
\end{align}
In this work we obtain the hydrodynamics of the vortices in the incompressible Euler fluid. However, our results are applicable to a  broader set of fluids which are approximately incompressible at distances away from vortices. The Gross-Pitaevsky equation in the regime where the vortex core is smaller than the separation between vortices is an example \cite{Lin}.

A convenient framework to develop the hydrodynamics of the vortex fluid is the Onsager's description  of  flows by a  system of a large but finite number of  point-like vortices. It is based on   Kirchhoff \cite{Kirchhoff} equations which we now recall (see e.g.,\cite{Kozlov}).

\smallskip

\paragraph*{2. Kirchhoff Equations as Lagrangian specification of the vortex flow.}
In two dimensions the curl of the  Euler equation  
\begin{align}\la{EE}
        D_t \bm u=-\bm \nabla p,\quad \bm\nabla\cdot  \bm u=0
\end{align}
for an   incompressible
fluid with a constant
density yields the single (pseudo) scalar equation for the vorticity 
\(\omega\equiv\bm\nabla\times  \bm u=\p_x u_y-\p_y u_x =2i \bar \p u \) (sometimes called Helmholtz equation)
\begin{align}
 \la{E}D_t 
        \omega=0.
\end{align}

In this form the Euler equation  has a simple geometrical meaning: the material derivative  \(D_t \equiv(\p_t+\bm{u}\cdot\bm\nabla)\) 
of the vorticity vanishes. Vorticity is transported along the divergence-free velocity field \(  \bm{u}\).

Helmholtz (1867), and  later Kirchhoff (1883) showed that there is a class
of solutions of the vorticity equation \eq{E} which consists of a finite
number of  point-like vortices. In this solution the
complex velocity of the fluid $u$  is given by a rational function    
\begin{align}
 \label{k}
        u (z,t)=-i \sum^N_{i=1 }\frac{\gamma_i}{z-z_i(t)} \,.
\end{align}
The  number of  vortices (poles) $N$  and the circulations (residues)   
\(2\pi\gamma_{i}\)  do not change in time,   while the moving positions
of vortices \(z_i(t)\) obey the  Kirchhoff equations:
\begin{align}
 \label{9}
        \dot {\overline z}_i=-i \sum^N_{j,j\neq i}\frac{\gamma_j}{z_i(t)-z_j(t)}.
\end{align}  
Kirchhoff equations replace the non-linear PDE \eqref{E} by a dynamical
 system.   The equations describe chaotic motions of a finite number of
vortices if \(N>3\). If $N$ is large  equations  can    approximate virtually any flow \cite{M}.

 We  will consider a \emph{chiral flow}, where  all vortices have the  counterclockwise circulation \(\omega>0\) and focus on the simplest case where all circulations are equal 
 \(\gamma_i=\gamma>0\).   

Under these specifications the velocity of the fluid, the stream function \(\bm{u}=(\p_y\psi,-\p_x\psi)\), or \( u =2i \p\psi\),  and Kirchhoff equations read
\begin{align}
 \label{100}
        u =-\sum^N_{ j=1}\frac{i \,\gamma}{z-z_j(t)},
        &\quad
        \psi=-\gamma\sum_{j=1}^N\log|z-z_j|,
 \\
        \dot {\overline z}_i= v _i,\quad v _i&=-\sum^N_{j, j\neq i}\frac{i \,\gamma}{z_i(t)-z_j(t)}.
 \la{10}
\end{align}
We want to study this  vortex system in the limit of a large number of vortices
(\(N\to\infty) \) 
with a given mean density  \(\bar\rho\).  Such fluid on the average performs a solid rotation. The frequency  of the solid rotation is approximately \(\Omega\approx \pi\gamma \bar\rho \).
 The mean density introduces the length scale    \(\ell\sim \bar\rho^{-1/2}\) which measures the average distance between vortices. Before we proceed to the large \(N\) limit we  describe some basic facts about the Kirchhoff equations.

\smallskip

\paragraph*{3. Hamiltonian structure.}
Kirchhoff equations are Hamiltonian. Holomorphic \(z_i\) and anti-holomorphic \( \bar z_i\) coordinates of vortices are canonical coordinates  with Poisson's brackets 
\begin{align}
 \la{131}
        \{z_i,\,\bar z_j\}=i  (\pi\gamma)^{-1}\delta_{ij} \,.
\end{align} 
In the case of the chiral flow the Hamiltonian  is  
\begin{align}
 \la{12}
        \mathcal{H}=-2\pi\gamma^{2}\sum_{i<j}\log|z_i-z_j| \,.
\end{align}   
In the next sections we compare this Hamiltonian with the
energy of an ideal fluid 
\begin{align}
 \la{141}
        H=\frac 12\int \bm{u}^2 d^2r.\
\end{align}
They are different. Kirchhoff's Hamiltonian captures only the part of the energy of the fluid which is transported  by vortices.  

Velocity of  vortices and the Kirchhoff equations  \eq{10} follow from the
canonical structure (\ref{131}) and the Hamiltonian (\ref{12})
\begin{align}
 \la{151}
        v_i\equiv\{\mathcal{H},\bar z_i\}
        =\frac{i }{\pi\gamma}\frac{\p\mathcal{H}}{\p z_i}\,.
\end{align}  

 In Cartesian coordinates  \(\bm{v}_i=\frac{1}{2\pi\gamma}(\p_{y_i}\mathcal{H}, -\p_{x_i}\mathcal{H)} \). The velocity is the skew gradient of the Hamiltonian.

More generally the evolution of a field \(\mathcal{O}(r_1,\dots,r_N)\) which
depends on vortex positions  is given by
\begin{align}
        \mathcal{\dot O}=\{\mathcal{H,\mathcal{O}}\}
        =\sum_i(\bm{v}_i\cdot\bm{\nabla}_{r_i})\mathcal{O} \,.
\end{align}  
Naturally, the vortex coordinates appear as Lagrangian coordinates of the vortex fluid.

\smallskip

\paragraph*{4. Eulerian specification of the vortex flow.}
Kirchhoff equations provide the Lagrangian specification of the vortex flow. We want to proceed to the Eulerian specification. We consider flows where     the microscopic density of vortices can be treated as  a coarse-grained smooth density function, i.e.  
\begin{align}
 \la{13}
        \sum_{i}\delta(r-r_i)\to \rho(r) = (2\pi\gamma)^{-1}\omega(r) \,.
\end{align} 
 Then the Eulerian field  \(\mathcal{O}(r)\) corresponding to the symmetric function of vortex coordinates  \(\mathcal{O}(r_1,\dots,r_N)\) can be obtained as a result of coarse-graining as
\begin{align}
        \sum_{i}\delta(r-r_i)\mathcal{O}(r_i) \to \mathcal{O}(r)\rho(r)  \,. 
 \la{Or}
\end{align}
In particular, the Eulerian velocity \(\bm{v}(r)\) is defined through the vortex flux 
\begin{align}
 \la{14}
        \sum_{i}\delta(r-r_i)\bm{v}_i\to \rho(r)\bm{v}(r)\,,
\end{align} 
where the velocity \( v_i\) is given by  \eq{10}.

Once we defined the vortex flux  \(\rho v\), we can find the coarse-grained energy \eq{12}. We use \eq{151} to write $(\pi\gamma)\rho v =i \sum_i\delta(r-r_i)\p_{z_i}\mathcal{H}$ and use the formula  $\sum_i\delta(r-r_i)\bm\nabla_{r_i}=\rho\bm\nabla  \frac{\delta}{\delta\rho}$ which converts between symmetric function of Lagrangian coordinates $r_{i}$ and functionals of Eulerian density $\rho$. We obtain the relation
\begin{align}
 \la{15}
         v =\frac{i }{\pi\gamma}\p \frac{\delta\mathcal{H}}{\delta\rho}.
\end{align}
This formula shows that the vortex fluid is incompressible as the original liquid itself and that the stream function of the vortex flow defined as \( v =-2i \p\Psi\) is given by $\Psi=-(2\pi\gamma )^{-1}{\delta\mathcal{H}}/{\delta\rho}=-{\delta\mathcal{H}}/{\delta\omega}$.
In the rest of the paper we use the density of vortices $\rho$  and the vorticity $\omega$ interchangeably.
\smallskip

\paragraph*{5. Velocity of the vortex flow and  deflections of the vortex stream lines.}
We start by computing the velocity of the vortex fluid. We use the $\bar \p$ formula $\bar\p \left(\frac 1z\right)=\pi\delta(r)$, write the vortex flux as   $\rho v=\frac{1}{\pi}\bar\p\left(\sum_i\frac{v_i}{z-z_i}\right)$,  substitute \eq{10}, use the identity 
\begin{align}
  \la{18}
        2\sum_{i\neq j}\frac{1}{z-z_i}\frac{1}{z_i-z_j}
        =\left(\sum_i\frac{1}{z-z_i}\right)^2\!\!\! +\p\sum_i\frac{1}{z-z_i} 
\end{align}
and obtain the important relation between the velocities of these two flows 
\begin{align}
         v =u -2\eta i \p\log \omega =u +\frac{\eta}{\omega}\Delta u,
 \la{23}
\end{align} 
where $\eta$ is given by \eq{eta}. We comment that in this work  both $\eta$ and  $\omega$ are set positive, but the product \(\eta\omega\) is positive regardless of the convention.\

The last term in (\ref{23}) is the source of anomalous terms in the hydrodynamics of the vortex fluid. The anomalous terms  reflect the discreetness  of vortices. Formally, the source of the anomalous term is  the  careful treatment of  \(i\neq j\) term
in sums over vortices positions in (\ref{9},{10},{12}) when  passing to
the  Euler specification. This is  seen from the identity
\eq{18}. The anomalous terms represent the excluded volume
occupied by a vortex.   The following simple argument helps to determine the coefficient in front of the anomalous term.  The velocity of the fluid close to the vortex core diverges as \(\gamma/z\).  This  pole singularity is cancelled by the anomalous term. As a result the velocity of the vortex  fluid (or the coarse-grained velocity of the fluid)  is a smooth function in contrast to the fluid  velocity. 
 The identity \eq{18} illustrates this effect: while each of two terms in the r.h.s. of \eq{18} has double poles, the  expression in the l.h.s. possesses only single poles.
 In the rest of the paper we trace the  consequences of the anomalous term.

  As an illustration of the anomalous effect let us use  \eq{23} to calculate the anomalous correction to the ``orbital moment'' of the vortex fluid $\int d^2r\,\rho [\bm{r}\times(\bm{v}-\bm{u})]=\eta\int d^2r\,(\bm{r}\cdot\bm{\nabla}\rho)=2\eta \int d^2r\,\rho$. We see that the correction is equal to $2\eta$ per vortex. This calculation is a continuous version of the elementary sum rule $\sum_i(\bm r_i\times \bm v_i)=2\eta N (N-1)$ for Kirchhoff vortices. It isolates the anomalous kinetic coefficient $\eta$.
 
The linearized relation of \eq{23} is 
$v \approx\left(1+\frac{1}{4}(\ell^2\Delta) \right)u$, 
where \(\ell=(2\pi\bar\rho)^{-1/2}\). 
Consider, for example a shear flow. If the flow is  oscillatory, say \(\bm{u}\sim (\cos{ky},0)\) the vortex flow lags the fluid \(v=(1-\frac 14(k\ell)^2)u\).
 However,
if the velocity of the  flow falls or raises, say as \(\bm{u}\sim(e^{-ky},0)\),
then the vortex flow \(v=(1+\frac 14(k\ell)^2)u\) leads the fluid. 

Eq.\eq{23} shows that the vortex flow is incompressible as the fluid, but the stream lines of both fluids deflect    
\begin{align}
 \la{27}
        \Psi=\psi-\eta\log|\Delta\psi| \,.
\end{align}
If we treat  the vortex flow as a coarse-grained flow we conclude  that the coarse graining causes a deflection of stream lines.

Yet another effect is a difference between the vorticities of two flows
\begin{align}
 \la{280}
        \bm\nabla\times \bm{v}=\omega+\eta\Delta\log \omega \, .
\end{align} 
We observe that the relation between vorticities is nonlinear. This causes interesting effects. 

If, for example, a  flow is a  large vortex patch performing a solid rotation $\bm{v}=\bm\Omega\times \bm r$ (the angular velocity is
a vector directed normal to the plane) the relation \eq{280}  inside the patch becomes   the  non-linear Liouville-like equation  
\begin{align}\la{KK}
         2\Omega=\omega+\eta\Delta\log \omega \,.
\end{align}
Solution of this equation depends on the flow outside the patch and may not correspond to a solid rotation of the fluid $\omega=2\Omega$.
For example, if  the patch is surrounded by the irrotational flow,  the equation \eq{KK} determines a 
 peculiar  structure of the interface. The interface   features   oscillations with a period of the order $\ell$ propagating well inside the patch. The oscillations are due to the anomalous term and reflect the discrete nature of vortices.

\smallskip

\paragraph*{6. The Hamiltonian of the vortex flow.} 
Once we know the vortex velocity, the relation \eq{15} helps to compute  the Hamiltonian  \begin{align}
 \la{235}
        \mathcal{H} =H-\eta\int \omega\, \log |\omega|\,  d^2r.
\end{align}
The anomalous term in \eq{235} (the difference between \(\mathcal{H}\) and \(H\)) is the energy of vortices at rest  \cite{comment3,comment4}. 

Using Eq.\eq{23} one can express Eq.\eq{235} in terms of velocity  of the vortex flow and the vortex density
\begin{align}
        \mathcal{H}=\frac 12 \int \left[\bm{v}^2 
        - \eta^2\left( \bm\nabla \log\rho\right)^2\right]d^2r \,.
\end{align}
The density of vortices is the only independent field in this flow. The Poisson algebra of the density field can be obtained directly from the Kirchhoff canonical brackets \eq{131} and the definition \eq{13}. They yield the familiar
brackets for the vorticity in incompressible fluid  \cite{1982-Olver}\begin{align*}
        \{\omega(r),\omega(r')\}\!=\!\frac {1}{2}(\bm\nabla_{r'} \times \bm\nabla_{r})
        [(\omega(r)\!+\!\omega(r'))\delta(r\!-\!r')] \,.
\end{align*}


\smallskip

\paragraph*{7. Momentum flux tensor and  pressure.}
We may read  the momentum flux tensor  from the Euler equation \eq{EE} written in  the form of the conservation law  \begin{align}
 \la{161}
        \dot u +\bar\p \Pi+\p\Pi_{z \bar  z}=0.
\end{align}
Here $\Pi$ and $\Pi_{z\bar z}$ are holomorphic component and the trace of the momentum flux tensor. Let us  recover
  the Euler equation  from   the Kirchhoff equations by computing the  evolution of velocity \(\dot u =\{ \mathcal{H},u\}\) with
the help of Kirchhoff's Hamiltonian structure.  We obtain  
\begin{align}
 \la{271}
        \Pi=-2 i \gamma\sum_i\frac{ v _i}{z-z_i},\quad \Pi_{z \bar z}
        =2 \gamma\,{\rm Im}\sum_i\frac{\bar v _i}{z-z_i}.
\end{align}
We observe the relation between the vortex flux and the traceless part of the momentum flux tensor
\be \la{250}
        (2\pi\gamma)\rho v =\omega v=i \bar\p\Pi\,,
\ee
and find it with the help  the  expression \eq{23} \begin{align}
 \la{19}
        &\Pi= u ^2- 4i\eta \p u.
\end{align}
 The last term is the anomaly. The stress exerted by the vortex fluid  differs from the
stress \(u ^2 \) exerted by the fluid itself by the anomalous stress  as in \eq{40}. The anomalous stress is a pseudo-tensor.  It represents a real force that could be measured. 

The origin of the anomalous term is the absence of vortex self-interaction in the sum \eq{12}. The coarse-grained stress  is a smooth function in contrast to the fluid  stress \(u ^2\).  One can see this from the following simple argument. The velocity of the fluid close to vortex core diverges as \(\gamma/z\). The stress of the fluid \(u^2\) diverges even stronger as \(\gamma^2/z^2\). The  second order pole singularity is cancelled by the anomalous term. 

The momentum flux tensor  \eq{271} can be written in Cartesian components as
\begin{align}
         &\Pi_{ab}=u_a u_b+\delta_{ab}\,(p+\eta\omega)+\tau_{ab}[u]\quad 
 \la{301}
\end{align}
and differs from   the fluid  momentum flux tensor \(u_a u_b+\delta_{ab}\,p\)  by the anomalous  tensor $\eta\omega\delta_{ab}+\tau_{ab}[u]\).  The second term here is given by \eq{2} with $v$ replaced by $u$. The anomalous correction represents forces exerted by vortices on the fluid.

The anomalous momentum flux  tensor is divergence-free $\nabla_b(\eta\omega\delta_{ab}+\tau_{ab}) =0$, and, therefore, does not contribute to the Euler equation of the fluid. Later we see that it contributes to the Euler equation \eq{36} of the vortex fluid. 

The  anomalous momentum flux tensor  \eq{301} has physical implications. Consider for example the Bernoulli law. It states that $2p+|u|^2$ stays constant along stream lines $\psi=\text{const}$ of a stationary flow of the fluid.  Instead, the trace of the momentum flux \eq{301} 
 \begin{align}
        \Pi_{z\bar z}=2p+|u|^2+2\eta  \omega=\text{const }
\end{align}
 does not change  along  stream lines \(\Psi={\rm\ const}\) of the vortex flow. These lines are different (see \eq{27}).

To conclude this paragraph we mention that the  general relation between the momentum flux tensor and  the energy  remain the same as in the Euler fluid 
\begin{align}
        \quad \bar\p\Pi=2\rho\p\frac{\delta\mathcal{H}}{\delta\rho}\,.
\end{align}

\smallskip

\paragraph*{8. Euler equation for the vortex flow.} 
In this paragraph we obtain the analog of the  Euler equation for the vortex flow. 
We treat vortices as elementary constituents of the vortex fluid which density is the vorticity  \(\rho=(2\pi\gamma)^{{-1}}\omega\)  and velocity \(\bm v\) defined by  (\ref{13},\ref{14}). Local expressions (\ref{23}-\ref{280}) warrant that the Euler equation for the vortex fluid is also local.  

The vortex flow is incompressible \(\bm\nabla\cdot \bm{v}=0\). 
Let us  introduce the material derivative of the vortex flow \(\mathcal{D}_t=\p_t+\bm{v}\cdot\bm\nabla
\).  Then the Helmholtz equation \eq{E} states that the material derivative of the vortex density vanishes 
\begin{align}
        \mathcal{D}_t\rho=0.
\end{align}   

The calculation of the material derivative of the vortex velocity is straightforward 
albeit tedious. It follows from  the relation between the vortex velocity and the fluid velocity \eq{23} and the evolution of the fluid velocity \eq{EE}. The result is 
\begin{align}
 \la{36}
        & \mathcal{D}_t  v_a+\nabla_a\mathfrak{p}+\rho^{-1}\nabla_b[\rho\tau_{ab}]=0 \,, 
\end{align}
where  \(\tau_{ab}\) is the anomalous  stress given by  (\ref{2}). The pressure \(\mathfrak{p}\)  can be found from the condition of incompressibility of the vortex fluid.
It is given by \eq{p} and also has an anomalous correction if compared to the pressure of the fluid $p$.

The anomalous stress  exerted by the vortex fluid  must be compared with the Coriolis force  \( -2\bm\Omega\times
\bm{v}\), where   \(\bm\Omega \) is  the angular velocity of the solid rotation.  The shift
of velocity in \eq{23} can be seen as a local transformation of coordinates
adjusted to a shear flow. Then the anomalous momentum flux tensor in \eq{36} and the anomalous corrections to pressure \eq{p} are the analogs
of the Coriolis force and the
centrifugal force, respectively.  

\smallskip

\paragraph*{9. Vortex flow on a curved surface: accumulation
of vortices at patches of positive curvature.} 
Anomalous terms are revealed on a curved surface. Here we describe only one effect: accumulation/deficit  of vortices in regions of space with positive/negative curvature. 
A comment is in order. A generalization of the vortex dynamics to the curved space requires the information of the energy of each vortex at rest in the curved background. This energy is additive and was ignored in the flat background. In the curved space it  depends on the metric, not necessarily in a universal manner, which causes an additional drift  of a single vortex in the presence of curvature gradients. We continue to omit this effect emphasizing the interaction between vortices.

Let us examine how the relation between the stream functions of the fluid and of the vortex fluid  \eq{27} changes when the Eucledian metric is deformed into the Riemanian metric. Under this deformation  the density of vortices is transformed as  the inverse form
of volume $\omega\to \omega\sqrt{g}$. Then the difference between stream functions transforms as $\Psi-\psi\to \Psi-\psi-\eta\log\sqrt g$. The anomalous transformation changes the relation between  vorticities of two fluids \eq{280}. The latter becomes
\begin{align}
 \la{531}
       2\Omega=\bm\nabla\times \bm{v}=  \omega+\eta\left(\Delta_g\log|\omega|
        - R\right)\,,
\end{align} 
where $\Delta_{g}$ is the Laplace-Beltrami operator on a space with metric $g$ and curvature $R$  and $\Omega$ is an angular velocity of the local rotation of the vortex fluid.

In the   leading order of the gradients   \eq{531} yields the transformation of the density of vortices at a given vorticity of the vortex flow $\Omega$
\begin{align}
 	\delta\omega=\eta\left[\delta R-\frac{1}{4}(\ell^2\Delta_g) 
	\delta R+\dots\right]
\end{align} 
In particular a conical singularity with a deficit angle \(\theta
\) accumulates an additional  \(2\eta\theta  \) vortices due to the anomalous forces. 

This result gives an alternative definition of the anomalous kinetic coefficient
\eq{eta}. It could be seen as a response of the vorticity to a variation
of the curvature $\eta=\delta\omega/\delta R$.

The formula establishes a global relation between the total vorticity of the fluid   and the vorticity of the vortex fluid.  We obtain it by  integrating \eq{531} over
the volume and using the Gauss-Bonnet formula.  We have 
\begin{equation}
        \int( \omega-\bm\nabla\times \bm{v})\, dV=4\pi\eta\chi \,.
  \la{euler-char}
\end{equation}

We mention one  particular application of this formula: if the vortex patch with the total vorticity \(2\pi\gamma N\) performs a solid rotation with a frequency \(\Omega\), the patch occupies the area which is less than  \(\pi\gamma N/\Omega\) by the anomalous amount $2\pi\eta\chi/\Omega^{}$. This result  illustrates the effect of the anomaly:
a local regularization  at a micro-scale yields effects on a macro-scale
and eventually   affects the global  relations. 
 
The most interesting effects of the anomalous terms are seen at the boundary.
We will discuss the boundary problem elsewhere.

We are grateful to I. Rushkin adn E. Bettelheim for  inputs at different stages of this work.
P.W. acknowledges  useful discussions  with G. Falkovich, G. Volovik, and T. Can and  thanks the International Institute of Physics (Brazil) and Weizmann Institute of Science (Israel) for the hospitality during the completion of the paper.  The work of P.W. was supported by NSF DMS-1206648, DMR-DMS-1206648, BSF-2010345 and by John Templeton Foundation. The work of A.G.A. was supported by the NSF under grant no. DMR-1206790.



\begin{thebibliography}{10}
\bibitem{comment1}A quest for the hydrodynamics of vortex fluid is found
in early works on
hydrodynamics  of superfluids \cite{Feynman,
Khalatnikov}.
\bibitem{AF} Anomalous forces acting on vortices in various settings are known in  superconductors and  superfluid $^3 He$,  see, e.g., \cite{Kopnin},   dense quark-gluon matter, see e.g.,\cite{Son1}, and have been recently discussed
 in connection to the Quantum Hall Effect \cite{Avron,W2,Tokatly,ReadHV,Son,Abanov, W2}
\bibitem{comment4} The same Hamiltonian has been obtained in \cite{WZ} in
studies of the 2D Dyson gas.
\end{thebibliography}
\end{document}